# On the Complexity of Exact Maximum-Likelihood Decoding for Asymptotically Good Low Density Parity Check Codes

Weiyu Xu and Babak Hassibi
EE Department
California Institute of Technology
Pasadena, CA 91125
weiyu,hassibi@caltech.edu

*Abstract*—Since the classical work of Berlekamp, McEliece and van Tilborg, it is well known that the problem of exact maximum-likelihood (ML) decoding of general linear codes is NP-hard. In this paper, we show that exact ML decoding of a classss of asymptotically good error correcting codes—expander codes, a special case of low density parity check (LDPC) codes—over binary symmetric channels (BSCs) is possible with an expected polynomial complexity. More precisely, for any bit-flipping probability, $p$, in a nontrivial range, there exists a rate region of non-zero support and a family of asymptotically good codes, whose error probability decays exponentially in coding length $n$, for which ML decoding is feasible in expected polynomial time. Furthermore, as $p$ approaches zero, this rate region approaches the channel capacity region. The result is based on the existence of polynomial-time suboptimal decoding algorithms that provide an ML certificate and the ability to compute the probability that the suboptimal decoder yields the ML solution. One such ML certificate decoder is the LP decoder of Feldman; we also propose a more efficient $O(n^2)$ algorithm based on the work of Sipser and Spielman and the Ford-Fulkerson algorithm. The results can be extended to AWGN channels and suggest that it may be feasible to eliminate the error floor phenomenon associated with message-passage decoding of LDPC codes in the high SNR regime. Finally, we observe that the argument of Berlekamp, McEliece and van Tilborg can be used to show that ML decoding of the considered class of codes constructed from LDPC codes with regular left degree, of which the considered expander codes are a special case, remains NP-hard; thus giving an interesting contrast between the worst-case and expected complexities.

## I. Introduction

ML decoding is a central algorithmic problem in coding theory[1], [2] since ML decoders minimize the message error probability when each codeword is transmitted with equal probability. For general linear block codes over binary symmetric channels (BSCs) the problem is as follows: given an $n \times m$ matrix $\mathbf{H}$ over $F_2$, a target vector $\mathbf{y} \in F_2^m$, and an integer $w > 0$, is there a vector $\mathbf{v} \in F_2^n$ of weight $\leq w$, such that $\mathbf{v}^t \mathbf{H} = \mathbf{y}^t$? Berlekamp, McEliece, and van Tilborg [1] have shown that this problem is NP-hard using a reduction from the 3-dimensional matching problem.

Since the publication of [1], the computational complexity of ML decoding of general linear codes has been extensively studied. To name a few, Bruck and Naor [3] and Lobstein [4] have shown that the problem remains NP-hard even if the code is known in advance, and can be preprocessed as long as desired. Quite recently, Guruswami and Vardy [2] have shown that ML decoding problem of Reed-Solomon Codes is NP-hard, the first hardness result for a specific family of codes with non-trivial algebraic architecture.

To quote further from [2], "...there is no nontrivial useful family of codes for which a polynomial-time maximum-likelihood decoding algorithm is known (such a result would, in fact, be regarded a breakthrough)". The existing results are either for codes which are not asymptotically good or apply to too general a class of codes. (For $i = 1, 2, ...$, let $C_i$ be an $(n_i, k_i, d_i)$ linear code over $F_2$. The infinite sequence of codes $C_1, C_2, ...$ is said to be asymptotically good if $n_i \to \infty$, $k_i/n_i \to R$ and $d_i/n_i \to \delta$ for some nonzero $R$ and $\delta$.) Furthermore, in many applications worst-case complexity results may not be very useful. To quote from [5]: "Although we have, by now, accumulated a considerable amount of results on the hardness of ML decoding, the broad worst-case nature of these results is still somewhat unsatisfactory....Thus it would be worthwhile to establish the hardness of ML decoding in the average sense, or for more narrow classes of codes". The current paper deals with both these issues; we show that a certain class of asymptotically good LDPCs (so-called expander codes) admit average polynomial-time ML decoding over BSCs and binary input AWGN (BI-AWGN) channels in certain rate regions.

This result is true regardless of whether preprocessing is applied to the code or not (Note without preprocessing,the expansion property of the Tanner graph is not available to the decoder in advance).It is based on the existence of polynomial-time suboptimal decoders with an ML certificate property. By this we mean that, in some cases, the decoder can certify that the solution is the ML solution; thus, we either get an exact ML codeword or declare an error. Let the code length be $n$, the rate $r$, the complexity of the suboptimal decoder $N(n)$, and the probability that the suboptimal decoder does not find the ML solution be $P_e(n)$. Then if we perform exhaustive search over the codebook whenever the suboptimal decoder fails to give an ML certificate, the expected complexity of the resulting ML algorithm, $EN_{ML}(n)$, will clearly be

$$EN_{ML}(n) = N(n) + k(n)P_e(n)2^{nR}, \qquad (1)$$

where $k(n)$ is a polynomial constant representing the compu-tation incurred per codeword in the exhaustive search. Clearly, if $P_e(n)2^{nR} \to 0$, then the ML algorithm will have expected polynomial-time complexity (equal to the complexity of the suboptimal decoder). Therefore in the remainder of the paper the main effort is to determine ML certificate decoders and to compute (or bound) $P_e(n)$.

One such ML certificate decoder is the LP decoder of Feldman [11]. We also propose a more efficient ML certificate decoder (reducing the worst-case complexity of the LP decoder from $O(n^9)$ to $O(n^2)$) which is based on the work of Sipser and Spielman [12] and the Ford-Fulkerson algorithm. We further characterize the achievable rate region $R_{ML}$ (albeit loosely) in which there exists a family of asymptotically good expander codes whose error probability goes to zero exponentially under an exact ML decoding algorithm with expected polynomial complexity. Finally, we observe that the argument of Berlekamp, McEliece and van Tilborg [1] can be used to show that exact ML decoding of the considered class of codes constructed from LDPC codes with regular left degree, of which the considered expander codes are a special case, remains NP-hard. This is reminiscent of the Hamilton path problem in graph theory which is NP-hard in the worst case but has an algorithm with average-case polynomial-time complexity [6].

We would like to point out that the results could have practical implications for eliminating the error floor phenomenon associated with message-passing decoding of LDPC codes in the high SNR regime [7], [8], [9], [10]. Although the exact ML algorithms in this paper are not practical by themselves, they may provide new insight into improvements of suboptimal message passing algorithm in this regime.

This paper is organized as follows. Section II studies exact ML decoding for expander codes with expected polynomial complexity under unlimited preprocessing of the codes. Section III proposes a new ML certificate algorithm and shows that exact ML decoding in expected polynomial-time is possible even without preprocessing. The rate region $R_{ML}$ is also characterized in Section III. Section IV gives the brief proof of the NP-hardness of ML decoding of the considered class of codes constructed from LDPC codes with regular left degree.

## II. EXPECTED POLYNOMIAL COMPLEXITY EXACT ML DECODING WITH UNLIMITED PREPROCESSING

We begin by allowing for unlimited preprocessing of the codebook as it makes the problem simpler and sets the stage for the subsequent proofs. Thus, consider a BSC with bit flipping probability $0 < p < \frac{1}{2}$. A Tanner graph $G$ is called a $(k, \Delta)$-expander if for every set $S$ of variable nodes where $|S| \leq k$, the number of check nodes incident to $S$ is larger than $\Delta|S|$[12]. We consider the family of binary parity-check left-regular expander graphs $G$ and its corresponding binary code $C'$ [12] with $n$ coded bits, $m$ parity checks and rate at least $(1 - m/n)$. Throughout the paper, $G$ has regular left degree $c$ for the variable nodes. We change the code $C'$ by adding more parity check constraints to make a new code $C$ with length $n$ and rate $r \leq (1 - m/n)$ when there exist dependent parity checks in $G$. Since $C$ is a subcode of $C'$, we have

*Lemma 1:* The minimum distance $\mathbf{w}_{min}$ of the new code $C$ of rate $r$ is no smaller than the minimum distance of the code $C'$.

Pick a random codeword in $C$ for transmission and denote the received sequence by **r**. Here is an ML certificate algorithm, where $d_H(\cdot, \cdot)$ denotes the Hamming distance.

- **Preprocessing**
    1) Compute the minimum distance of the code $C$
- **Decoding**
    1) If there is a variable that is in more unsatisfied than satisfied constraints (only the constraints in the expander graph $G$), flip the value of that variable
    2) Repeat 1) until no such variable remains. Denote the resulting sequence as $\mathbf{x}'$;
    3) If $\mathbf{x}'$ is in the code $C$ and $d_H(\mathbf{x}', \mathbf{r}) \leq \frac{\mathbf{w}_{min}}{2}$ declare $\mathbf{x}'$ as the ML sequence. Otherwise, declare an error.

Step 1) and 2) in **Decoding** is basically the "simple sequential decoding" algorithm of [12]. The novelty here is that, if the minimum distance is known, an ML certificate can be provided. If we perform exhaustive search whenever the algorithm fails, we have an ML algorithm. To compute the probability of failure, $P_e(n)$, we need a lemma on the minimum relative distance (defined as $\mathbf{w}_{min}/n$) of $C$.

*Lemma 2:* Let the bipartite expander graph $G$ be a $(\alpha n, 3c/4)$ expander, where $0 < \alpha < 1$ and $c$ is the constant left-degree. Then the minimum relative distance of $C$ is at least $3\alpha/2$.

*Proof:* Suppose that the minimum relative distance for the code $C'$ is $\beta < 3\alpha/2$. Then the set **S** of variable nodes having value '1' in a non-zero codeword with relative weight $\beta$ is connected to at least $\frac{2}{3} \times \beta n \times (3c/4) > (c/2)\beta n$ parity check neighbors. Since each variable node can accommodate at most $c$ check nodes, there must be at lest one check node neighboring **S** that is connected to only one variable node in **S**. But then that parity check can not be satisfied, a contradiction. Combining $\beta \geq 3\alpha/2$ with Lemma 1, we get Lemma 2. ∎

**Note**: In Lemma 2, we assume $\alpha n, \beta n$ and $2\beta n/3$ are all integers, which for large enough $n$ is not an issue.

We can now compute $P_e(n)$.

*Lemma 3:* Let bipartite Tanner graph $G$ be an $(\alpha n, 3c/4)$ expander. Then $P_e(n)$ is upper bounded by by $2^{-D(\frac{\alpha}{2}\|p)n}$, where $\frac{\alpha}{2} > p$ and $D(x\|y) = x \log_2 \frac{x}{y} + (1 - x) \log_2 \frac{1-x}{1-y}$ is the Kullback-Leibler divergence between Bernoulli random variables with parameters $x$ and $y$ respectively.

*Proof:* For the expander code $C'$ corresponding to $G$, the simple sequential decoding algorithm can correct up to $\frac{\alpha}{2}$ fraction of errors (Theorem 10 in [12]). Since $C$ is a subcode of $C'$, the same is true of $C$. By Lemma 2, the minimum relative distance for $C$ is at least $\frac{3\alpha}{2}$. So if the fraction of errors is no larger than $\frac{\alpha}{2}$, the algorithm will provide the ML certificate. For $p < \frac{\alpha}{2}$, the probability of having no less than $\alpha/2$ fraction of errors occurring is upper bounded by $2^{-D(\frac{\alpha}{2}\|p)n}$ from a standard Chernoff bound argument. ∎

*Lemma 4:* For a fixed $0 < x < 1$, $D(x\|y)$ goes to infinity as $0 < y < 1$ approaches 0.

*Proof:* Straightforward computation. ∎

We can now give the main result of the section.

*Theorem 1:* Let Tanner graph $G$ be an $(\alpha n, 3c/4)$ expander. Then the expected computational time of an ML decoding algorithm over the BSC channel, $N_{ML}(n)$, is polynomial in $n$, given that $p < \frac{\alpha}{2}$ and $r < D(\frac{\alpha}{2}\|p)$. The storage complexity is kept polynomial in $n$ in the worst case.

*Proof:* The first statement follows from Lemma 3 and (1). For the second statement, we note that since each codeword is generated one by one the storage complexity is kept polynomial in $n$ in the worst case. ∎

Rather than fixing $p$ and looking at the rate, we can fix the rate $r$ and look at the flipping probability $p$.

*Theorem 2:* For any rate $0 < r < 1$, there exists a threshold $0 < p^* < 1$ and a family of asymptotically good block codes with rate $r$ and length $n$ such that when allowing preprocessing of the codes, exact ML decoding can be achieved with an expected polynomial complexity in $n$ over the BSC with bit flipping probability $0 \leq p < p^*$. Furthermore, the block error probability of this family of codes decreases exponentially in $n$ asymptotically.

*Proof:* For any rate $r$, with sufficiently large but constant left degree $c$ and right degree $d$ ($r = 1 - \frac{c}{d}$), there exists a number $\alpha > 0$ and a family of expander graphs $(\alpha n, 3c/4)$ with the number of variable nodes $n$ going to infinity [14]. By Lemma 2, such expander graphs give a family of codes of length $n$ with minimum relative distance of $3\alpha/2$ and rate $r$. Since $\frac{\alpha}{2} > 0$, by Lemma 4, there must be a number $p^*$ so that for any $p < p^*$, we have $p < \frac{\alpha}{2}$ and $r < D(\frac{\alpha}{2}\|p)$. By Theorem 1, the first statement holds. For the second statement, since the ML decoder corrects up to $3\alpha/4$ fraction of errors, which is larger than $p$, we have an exponentially decreasing error probability from the Chernoff bound. ∎

*Lemma 5:* For any family of $(\alpha n, 3c/4)$ expander codes with increasing code length $n$ and a constant $\alpha > 0$, the improvement in the lower bound of the error exponent by using the ML algorithm instead of the "simple sequential decoding algorithm" [12] for this family of codes is arbitrarily large if $p$ is sufficiently small but remains positive. However, this improvement comes with expected polynomial complexity in $n$ when allowing preprocessing of the codes.

*Proof:* By using the Chernoff bound, the block error rate of ML decoding is upper bounded by $2^{-D(\frac{3\alpha}{4}\|p)n}$, while the block error rate of iterative decoding is upper bounded by $2^{-D(\frac{\alpha}{2}\|p)n}$ when $p < \frac{\alpha}{2}$. Since $D(\frac{3\alpha}{4}\|p) - D(\frac{\alpha}{2}\|p) \geq \frac{\alpha}{4}\log_2(\frac{3\alpha}{4}/p) + c'$, where $c'$ is a constant, this difference grows to infinity as $p \to 0$. ∎

We now briefly consider BI-AWGN channels while allowing preprocessings of the codes. To do ML decoding, we first make a hard decision on the received sequence $\mathbf{r}'$. Then we send the hard-decided sequence $\mathbf{r}$ to an exact ML decoder as in BSC channels except we change the Hamming distance to the Euclidean distance. Using a union bound, the $P_e(n)$ is upper bounded by $P_1 + P_2$, where $P_1$ is the probability of more than $\alpha/2$ bit errors occurring in $\mathbf{r}$ and $P_2$ is the probability that the noise vector's $l_2$ norm is larger than one half of the minimum Euclidean distance between codewords. By the Chernoff bound for Bernoulli and chi-square random variables (details omitted) and Lemma 2, we have $P_1 \leq 2^{-D(\frac{\alpha}{2}\|P_{err})n}$ and $P_2 \leq e^{-\frac{1}{2}(\frac{3}{2}SNR - 1 - \ln(\frac{3}{2}SNR))n}$, where $P_{err}$ is the hard decision bit error probability for BI-AWGN channels, which improves as the SNR increases. Using the same arguments as for the BSC, we have counterparts to Theorems 1 and 2 for BI-AWGN channels when the SNR is high enough.

In this section, the key to obtaining a polynomial-time ML certificate algorithm was knowing the minimum distance of the code (which must be precomputed). We now show that such ML certificate algorithms can be obtained without preprocessing.

## III. Expected Polynomial Complexity Exact ML Decoding without Preprocessing

In LP decoding over memoryless channels the problem is relaxed to a linear programming problem with polynomial complexity [11]. The LP decoder has the ML certificate property: if the solution to the relaxed LP is integral it is the ML codeword. The reader is referred to [11] for further details. In [13], by constructing a feasible point for the dual of the LP, the authors prove that the LP decoder can correct a constant fraction of errors when applied to expander codes of sufficient expansion, which is stated in the following theorem:

*Theorem 3:* [13] Let $C$ be a low-density parity-check code with length $n$ and rate at least $1 - m/n$ described by a Tanner graph $G$ with $n$ variable nodes, $m$ check nodes, and regular left degree $c$. Suppose $G$ is an $(\alpha n, \delta c)$-expander, where $\delta > 2/3 + 1/(3c)$ and $\delta c$ is an integer. Then the LP decoder succeeds and gives the ML solution, as long as at most $\frac{3\delta - 2}{2\delta - 1}(\alpha n - 1)$ bits are flipped by the channel.

The above theorem implies that $P_e(n)$ can be computed as in the previous section and that the LP decoder, in conjunction with exhaustive search, is an expected polynomial time ML decoder. However, the main disadvantage of LP decoding is the complexity coming from solving a linear program. As noted in [11], the worst-case total number of variables and constraints in the LP relaxation is of order $O(n^3)$ if we consider an expander graph with irregular check degrees, where $n$ is the length of the codes. To solve a general LP, the complexity is of order $O(dim^3)$, where $dim$ is the number of variables and constraints. This implies a complexity of order $O(n^9)$. We now show that the ML certificate property can be achieved with worst-case complexity $O(n^2)$, without sacrificing the guaranteed performance.

**New ML Certificate Algorithm**

1) If there is a variable that is in more unsatisfied than satisfied constraints (only the constraints in the expander graph), flip the value of that variable
2) Repeat 1) until no such variable remains. Denote the resulting sequence as $\mathbf{x}'$.
3) If $\mathbf{x}'$ is not in the code $C$, go to 4). Otherwise, construct a series of max-flow instances as follows: Introduce a source node $s$ and a sink node $t$. Let $U$ denote the set of variables nodes where $\mathbf{r}$ and $\mathbf{x}'$ differ, $N(U)$ denote the check node neighborhood of $U$ and let $\tilde{U}$ denote the set of variable nodes other than $U$ that are connected to

$N(U)$. Take an integer $A = \lfloor \frac{c}{2} \rfloor + 1$. Add directed edges from $s$ to the set of variables $U + \tilde{U}$. If $i \in U$, assign capacity $A$ to the directed edge from $s$ to $i$. For each $i \in \tilde{U}$, if $|N(i) \bigcap N(U)| > (2A - c)$, assign integer capacity $|N(i) \bigcap N(U)| - (2A - c)$ to the directed edge from $s$ to $i$, otherwise assign capacity 0 to it. Construct directed edges from any $i \in U \bigcup \tilde{U}$ to its neighbors in $N(U)$ and assign integer capacity 1 to them. Construct directed edges from each check node in $N(U)$ to the sink node $t$ and assign capacity 1 to them.

Use the Ford-Fulkerson algorithm to find the max-flow from the source $s$ to the sink $t$. If the max-flow value is equal to the sum of the capacities of the edges from $s$ to $U \bigcup \tilde{U}$, then declare $\mathbf{x}'$ as the ML sequence. If this does not hold but $A < c$, increase $A$ by 1 and construct a new max-flow instance. Otherwise, go to 4).

4) Declare an error.

When the capacities are integers the runtime of Ford-Fulkerson algorithm is bounded by $O(E * f)$, where $E$ is the number of edges in the graph and $f$ is the maximum flow in the graph [17]. $E$ is of order $O(n)$ since each variable node has a constant degree $c$ and $f$ is also of order $O(n)$ since it is upper bounded by $cn$. Obviously the max-flow algorithms are performed at most $c$ times. Since each step of the new ML certificate algorithm is of order $O(n^2)$, the total complexity is of order $O(n^2)$.

*Lemma 6:* If the max-flow attains the sum of edge capacities from the source $s$ to $U \bigcup \tilde{U}$ for some $A \geq \lfloor \frac{c}{2} \rfloor + 1$, then $\mathbf{x}'$ must be an exact ML codeword.

*Proof:* There is a maximum flow such that the flow through every edge is integral[17]. If the max-flow value is equal to the sum of edge capacities from $s$ to $U \bigcup \tilde{U}$, each node $i$ in $U$ is connected to a set $M(i)$ of $A$ nodes and $M(i) \bigcap M(j) = \phi$ if $i \neq j$. Suppose there is a codeword $\mathbf{x}''$ such that $d_H(\mathbf{x}'', \mathbf{r}) < d_H(\mathbf{x}', \mathbf{r})$. Then $\mathbf{x}''$ must share the same values with $\mathbf{r}$ in $l > 0$ positions of $U$(otherwise, $d_H(\mathbf{x}'', \mathbf{r}) \geq d_H(\mathbf{x}', \mathbf{r})$). We show that outside $U$, $\mathbf{x}''$ must be different from $\mathbf{r}$ in at least $l$ places, which contradicts $d_H(\mathbf{x}'', \mathbf{r}) < d_H(\mathbf{x}', \mathbf{r})$. Without loss of generality, we assume that $\mathbf{x}''$ is the all-zero codeword. Then the $l$ shared bits are $l$ '1's. In total, the shared bits in $U$ have at least $A \times l$ neighbor check nodes, in each of which there must be at least 2 variable nodes of value '1' to make $\mathbf{x}''$ a codeword. However, the $l$ shared bits can at most provide extra $(c - A) \times l$ '1's to fill the extra '1's, which is at least $2A \times l - A \times l = A \times l$, to satisfy the the check nodes in $N(U)$. So we still need $A \times l - (c - A) \times l = (2A - c) \times l$ '1's. But from the construction of the max-flow instance, each of the variable nodes outside $U$ can contribute at most $(2A - c)$ '1' to these check nodes. So there must be at least $l$ of variable nodes with value '1' outside $U$.

Although this new algorithm uses max-flow arguments as in the analysis of the LP decoder [13], there are several key differences. The max-flow argument in [13] is for the purpose of *analysis* in proving the existence of a dual feasible point, but in the new algorithm max-flow arguments are used *directly* in the *computation*, thus reducing the complexity by avoiding solving a large linear program. The *series* of max-flow instances here are much more refined because they use *optimized* rather than *uniform* link capacities from the source to the variable nodes in the max-flow instance of [13]. Moreover, without looking for the dual feasible edge weight assignment, the new algorithm and its *direct* proof provide more intuition about why expander codes efficiently correct a constant fraction of errors while having the ML certificate property even without preprocessing. The Ford-Fulkerson algorithm can be easily integrated into any belief propagation decoder to efficiently offer them the ML certificate property,which can help the decoder decide whether it is necessary to perform more computations to improve the performance. The following lemma gives a performance guarantee of the new ML certificate algorithm.

*Lemma 7:* Suppose $\delta > 2/3 + 1/(3c)$, and $\delta c$ is an integer. Then if $\mathbf{x}'$ is in the code $C$ and is different from $\mathbf{r}$ in at most $\frac{3\delta - 2}{2\delta - 1}(\alpha n - 1)$ positions, $\mathbf{x}'$ will be certified to be an ML codeword in the new ML certificate algorithm. When $\delta = 3/4$, the new algorithm is guaranteed to correct up to $\alpha/2$ fraction of errors while providing ML certificate property,which matches the proved capability of LP decoder.

*Proof:* In [13] the authors showed the conditions in this lemma imply the existence of a $\delta$-matching of $U$ (see the definition in [13]). We show that the number of edges connected to each variable node $i$ in a $\delta$-matching of $U$ is no smaller than the capacity from the source $s$ to $i$ for $A = \delta c$, thus implying a maximum flow as specified in Lemma 6. As in [13], define $\dot{U} = \{i \in V : i \notin U, |N(i) \bigcap N(U)| \geq (1 - \lambda)c + 1\}$ and let $\lambda = 2(1 - \delta) + 1/c$. If $i \in U$, the number of connected edges in $\delta$-matching is equal to the the capacity $A$. If $i \notin U \bigcup \dot{U}$, the capacity from $s$ to $i$ is zero because $(1 - \lambda)c - (2A - c) = -1 < 0$. If the variable node $i \in \dot{U}$,let $B$ be the number of check nodes among $N(U)$ that $i$ is connected to by the $\delta$-matching. From the definition of $\delta$-matching, the node $i$ will be incident to at least $\lambda c$ edges in the $\delta$-matching. Suppose $B < |N(i) \bigcap N(U)| - (2A - c)$, $i$ is incident to less than $|N(i) \bigcap N(U)| - (2A - c) + |N(i) \bigcap \overline{N(U)}| = |N(i)| - (2A - c) = 2c(1 - \delta) < \lambda c$ edges in the $\delta$-matching, a contradiction. Thus there will be maximum flow satisfying Lemma 6 since we can assign the flow from the source to the edges in the $\delta$-matching. Since by the step 1) and step 2) we can correct up to $\alpha/2$ fraction of errors for an expander graph with $\delta = 3/4$. Also, the maximum flow instances provide the exact ML sequence.

From Lemma 6 and 7, we see that the new ML certificate algorithm corrects a constant fraction of errors with low complexity. Since the LP decoder and the newly proposed ML certificate decoder can correct a constant fraction of errors without preprocessing, we have a counterpart of Theorem 2 for the case of no preprocessing allowed.

For a BSC channel with fixed bit flipping probability $p$, let us denote $R_{ML}(p)$ as the set of rates $t$ in which there exists a family of asymptotically good codes whose error probability goes to zero exponentially in the coding length under an expected polynomial complexity exact ML decoding algorithm without preprocessing. We now give an achievable region of $R_{ML}(p)$, for $0 < p < 1/2$. Clearly, $R_{ML}(p)$ has the channel capacity $1 - H(p)$ as an upper bound.

*Lemma 8:* For a fixed $0 < p < 1/2$, the rate set $R(p) \subseteq R_{ML}(p)$, where $R(p)$ is the set of rates $t$ such that $t \leq r, p < \frac{3\delta-2}{2\delta-1}\alpha, t < D(\frac{3\delta-2}{2\delta-1}\alpha||p)$, where $\alpha = (2e^{\delta c+1}(\delta c/(1-r))^{(1-\delta)c})^{-\frac{1}{(1-\delta)c-1}}$, for some $r, c, \delta$ satisfying $0 < r < 1, c \in N, (2/3 + 1/(3c)) < \delta < 1, \delta c \in N, (1-\delta)c \geq 2$.

*Proof:* It can be shown using random graphs that for the $r, \delta$ and $c$ in Lemma 8, there is a bipartite graph $G$ for any $n$ variable nodes and $(1-r)n$ check nodes, which is an expander $(\alpha n, \delta c)$[13]. By constructing a linear code with rate $t$ from the expander graph (noticing that the rate of the code $t$ can be made smaller than $r$) and applying the expected polynomial-time ML decoders described, we get the desired result. ∎

*Lemma 9:* If $p$ is sufficiently close to zero but remains positive, the gap between the channel capacity and the supremum of the rate region $R_{ML}(p)$ is arbitrarily small.

*Proof:* Take any $0 < \epsilon < 1$, let $r = 1 - \epsilon$, $t = r$ and choose any $c$ and $\delta$ according to Lemma 8. Then from Lemma 8, there is a $p^*$ such that for $0 < p < p^*$, such that $t = r \in R_{ML}(p)$. But the channel capacity for any $p$ in the region $0 < p < p^*$ is at most 1, which is no bigger than $t + \epsilon$. ∎

## IV. EXPECTED COMPLEXITY VERSUS WORSE-CASE COMPLEXITY: A CONTRAST

In this part, we prove that for any $0 < t < 1$, $t \leq r < 1$, the exact ML decoding problem remains NP-hard for the family of codes $C$ of rate $R \leq t$ constructed by adding linear constraints to the $LDPC$ codes $C'$ of rate at least $r$ defined by Tanner graphs $G$ with regular left degree $c \geq 3$. This family of codes correspond to the codes we discussed in the previous two sections with the difference that we do not require the Tanner graph to be an expander graph.

Let the newly added linear constraints, the linear constraints corresponding to the check nodes in $G$ and the syndrome $\mathbf{y}$ be revealed to the decoder. We show that even in this more restricted case, the ML decoding problem is NP-hard.

*Proof:* Our proof essentially follows that of [1]. As long as the flipping probability $0 < p < 1$, the received codeword can be any binary sequence of length $n$. We reduce the $k$ dimensional matching problem (k-DM) to the ML decoding problem, where $k = c$. It is known that the k-dimensional matching problem is NP-hard for $k \geq 3$ [15][16]. The decision problem for the $c$-dimensional matching problem is as follows: given a subset $U \subseteq T \times T \cdots \times T$, where $T$ is a finite set and the elements of $U$ are c-tuples from the set $T$, determine whether there is a set $W \subseteq U$ such that $|W| = |T|$, and no two elements of $W$ agree in any coordinate.

Without loss of generality, we assume that the cardinality of $U$ is larger than $|T|$, otherwise the corresponding c-dimensional matching problem will be trivial. Just as in [1], for any such c-DM problem, encode the set $U$ of c-tuples into a $|U| \times c|T|$ incidence matrix $M$, in which each row corresponds to one of the c-tuples and has weight $c$, with each 1 corresponding to a component of the c-tuple. Sequentially repeat each row of the binary incidence matrix by $I = \max\{\lceil \frac{c|T|}{r|U|} \rceil, \lceil \frac{1}{t} \rceil\}$ times to create a new binary incidence matrix $M'$ of size $(I|U|) \times (c|T|)$, where every consecutive $I$ rows are the same. Since $|U| > |T|$ and $t$ is fixed, $I$ is upper-bounded by a constant. Let the code corresponding to $M'$ be $C'$. Let the added $(I-1)|U|$ linear constraints be the simple constraints which specify the $(I-1)|U|$ bits corresponding to the duplicate copies in $M$ of any row in $M'$ to be zero. Combine these $(I-1)|U|$ simple constraints with $M'$, we get a new parity check matrix $M''$ of dimension $I|U| \times ((I-1)|U| + c|T|)$. Thus the new code $C$ corresponding to the new parity check matrix $M''$ is a valid example in the considered family of codes. Take $\mathbf{y} = (0, 0, 0, \ldots 0, 1, 1, \ldots, 1)^T$, where $\mathbf{y}$ has $(I-1) \times |U|$ 0's and $c|T|$ 1's. Suppose we have a polynomial-time algorithm for the ML decoding of the considered family of codes, we just run the putative ML decoding algorithm with the parity check matrix as $M''$, the syndrome as $\mathbf{y}$, and $w = |T|$, we will know the answer to the $k$-dimensional matching problem. So the ML decoding of the considered family of codes is NP-hard in the worst case. ∎

## ACKNOWLEDGMENT

We thank Professor R.J.McEliece for helpful discussions.